\documentclass[%
superscriptaddress,
 amsmath,amssymb,
prf,
onecolumn,
]{revtex4-2}
\usepackage[export]{adjustbox}
\usepackage{graphicx}% Include figure filessn
\usepackage{color}% Include figure files
\usepackage{xcolor}% Include figure files
\usepackage{natbib}
\usepackage{wasysym}% Include figure files
\usepackage{dcolumn}% Align table columns on decimal point
\usepackage{bm}% bold math
\usepackage{hyperref}

\begin{document}
% Full title of the paper (Capitalized)
\title{Turbulent convection at very high Rayleigh numbers and the weakly nonlinear theory}

% MDPI internal command: Title for citation in the left column
%\TitleCitation{Title}

% Author Orchid ID: enter ID or remove command
%\newcommand{\orcidauthorA}{0000-0000-0000-000X} % Add \orcidA{} behind the author's name
%\newcommand{\orcidauthorB}{0000-0000-0000-000X} % Add \orcidB{} behind the author's name

% Authors, for the paper (add full first names)
\author{Katepalli R. Sreenivasan} 
\affiliation{Department of Physics, Department of Mechanical and Aerospace Engineering, and the Courant Institute of Mathematical Sciences, New York University}
\email{krs3@nyu.edu}
%$^{1,\ddagger,*}$\orcidA{}, and Joseph J. Niemela $^{2,\ddagger}$}
\author{Joseph J. Niemela}
\affiliation{The Abdus Salam International Center for Theoretical Physics}
\email{niemela@ictp.it}

%\longauthorlist{yes}

% MDPI internal command: Authors, for metadata in PDF
%\AuthorNames{Katepalli R. Sreenivasan and Joseph J. Niemela}

% MDPI internal command: Authors, for citation in the left column
%\AuthorCitation{Sreenivasan, K.R; Niemela, J.J.}
% If this is a Chicago style journal: Lastname, Firstname, Firstname Lastname, and Firstname Lastname.

% Affiliations / Addresses (Add [1] after \address if there is only one affiliation.)
%\address{%
%$^{1}$ \quad Department of Physics, Department of Mechanical and Aerospace Engineering, and the Courant Institute of Mathematical Sciences, New York University; krs3@nyu.edu\\
%$^{2}$ \quad The Abdus Salam International Center for Theoretical Physics; niemela@ictp.it}

% Contact information of the corresponding author
%\corres{Correspondence: krs3@nyu.edu; Tel. +1-347-410-4509.}

% Current address and/or shared authorship
%\firstnote{Current address: Affiliation 3.} 
%\secondnote{The authors contributed equally to this work.}
% The commands \thirdnote{} till \eighthnote{} are available for further notes

%\simplesumm{} % Simple summary

%\conference{} % An extended version of a conference paper

% Abstract (Do not insert blank lines, i.e. \\) 
\begin{abstract}
To provide insights into the challenging problem of turbulent convection, Jack Herring used a greatly truncated version of the complete Boussinesq equations containing only one horizontal wavenumber. In light of later observations of a robust large scale circulation sweeping through convecting enclosures at high Rayleigh numbers, it is perhaps not an implausible point of view from which to reexamine high-Rayleigh-number data. Here we compare past experimental data on convective heat transport at high Rayleigh numbers with predictions from Herring's model and, in fact, find excellent agreement. The model has only one unknown parameter compared to the two free parameters present in the lowest order least-squares power-law fit. We discuss why the underlying simplistic physical picture, meant to work at Rayleigh numbers slightly past the critical value of a few thousands, is consistent with the data, when the single free parameter in it is revised, over some eleven decades of the Rayleigh number---stretching from about a million to about $10^{17}$.
\end{abstract}
\maketitle

% Keywords
%\keyword{turbulent convection; heat transport; high-Rayleigh-number asymptote; ultimate state of convection} 

% The fields PACS, MSC, and JEL may be left empty or commented out if not applicable
%\PACS{J0101}
%\MSC{}
%\JEL{}

%%%%%%%%%%%%%%%%%%%%%%%%%%%%%%%%%%%%%%%%%%
%\begin{document}

%%%%%%%%%%%%%%%%%%%%%%%%%%%%%%%%%%%%%%%%%%

\section{Introduction}

Turbulent thermal convection is a grand problem: if we understand it well enough, we might shed some light on the dynamics of the Sun and sun-like stars, as well as the long-term evolution of terrestrial planets. Its importance to engineering applications such as the cooling of nuclear reactors is obvious and needs no special emphasis. The scientific paradigm of thermal convection is the so-called Rayleigh-B\'enard convection (RBC), in which a layer of viscous fluid between two smooth horizontal plates is heated at the bottom wall and cooled at the top. In practice, the fluid is constrained by side walls which are non-conducting, so the heat input to the bottom plate is communicated to the top plate entirely through the fluid layer. A theoretical analysis of the resulting fluid flow usually incorporates the so-called Boussinesq approximation \cite{tritton}. Allowing non-Boussinesq effects opens up the problem to many incompletely understood features \cite{tritton,jfm,pand1,pand2}. %\cite{,Krishnamurti,Herring66, Spiegel,Kraichnan,Busse78,Nature}  

We had earlier set out to measure the heat transfer due to turbulent convection in a 1-m tall, diameter-to-height aspect ratio $\Gamma= 1/2$, using cryogenic helium gas as the working fluid \cite{Nature}. The purpose of using low temperature helium gas was to enable extremely high Rayleigh numbers (up to $10^{17}$) to be attained, while keeping conditions nearly Boussinesq (see \cite{jfm} for a detailed assessment), and to take advantage of the thermal isolation of the apparatus afforded by the cryogenic environment. The details of the experiment and its execution are provided in \cite{Nature,jfm} and summarized in the following section. Our goal in this short paper is to compare the analysis of Herring \cite{Herring63,Herring66} with experimental results on heat transport measured in high-Rayleigh-number turbulent convection. While Herring's formula was derived for the immediate supercritical state of convection where nonlinearities are weak, we find that it agrees with experimental data over 11 decades of Rayleigh number---all in the turbulent state. 

A few background comments are in order on the traditional framework of RBC. The thermal driving of the flow is the temperature difference between the top and bottom plates, which is measured in terms of the Rayleigh number defined as
%\begin{adjustwidth}{-\extralength}{0cm}
\begin{equation}
Ra = \dfrac{g \alpha \Delta T H^3}{\nu \kappa},
\end{equation}
%\end{adjustwidth}
where $g$ is the acceleration due to gravity, $\alpha$ is the isobaric thermal expansion coefficient, $\Delta T$ is the temperature difference across a vertical fluid layer of height $H$, and $\nu$, $\kappa$ are the kinematic viscosity and thermal diffusivity of the fluid, respectively. The response of the flow is the heat transport across the fluid height, measured in terms of the so-called Nusselt number, $Nu$, which is the actual amount of heat transport effected by convection to that possible (for the same $\Delta T$), by thermal conduction alone. One could also include the dependence of $Nu$ on the Prandtl number, $Pr = \nu/\kappa$, so a fundamental problem of thermal convection is to determine the functional dependence of $Nu$ on $Ra$ and $Pr$. Another response of the flow is the Reynolds number of the convective motion, but we shall not consider it here.

As in most other turbulence problems (and for all many-body problems in 3D), RBC also cannot be solved fully from a theoretical point of view, so there are only scaling theories. There are two schools of thought. That due to
Malkus \cite{Malkus} and Spiegel \cite{Spiegel1} says that \begin{equation}
Nu = 0.073 Ra^{1/3},  
\end{equation}
with no dependence on $Pr$. The underlying physics is the so-called marginal stability of the top and bottom boundary layers (see \cite{Howard}). On the other hand, Kraichnan \cite{Kraichnan} argued that the boundary layers will become irrelevant at {\it very} high $Ra$ and obtained the explicit formula 
\begin{equation}
 Nu = C Pr^{-1/4} \lbrace Ra/log(Ra)^3\rbrace^{1/2}.   
\end{equation}
on the basis of an analysis of top and bottom boundary layers (which were particularly incompletely understood then). Even though Kraichnan made valiant efforts to obtain the  constant $C$, the details are tenuous and so nothing is lost, at least for our purposes, in regarding it as an unknown constant. See also \cite{Spiegel2,Chavanne} for similar predictions of the Rayleigh number dependence but they did not go into any boundary layer details.

Because Kraichnan's formula demands very high $Ra$, the $\frac{1}{2}$-power dependence on $Ra$ (with logarithmic corrections), including an explicit Prandtl number dependence, is thought to represent the ``ultimate state” or the ``asymptotic state” of RBC. As Spiegel \cite{Spiegel2} remarked, the difference between formulae (2) and (3) needs to be resolved because it would then suggest the correct physics that operates at high $Ra$. We will take up this thread shortly.

From a different perspective, the weakly nonlinear theory of Herring, as summarized by Busse \cite{Busse78}, with a long pedigree involving Lou Howard, Willem Malkus, Paul Roberts and Fritz Busse, gives
%\begin{adjustwidth}{-\extralength}{0cm}
\begin{equation}
Nu = D \lbrace Ra^{3/2}ln(Ra)^{3/2}\rbrace^{1/5},
\end{equation}
%\end{adjustwidth}
where the constant $D = 0.24$ was theoretically calculated by maximizing the heat transport accomplished by a single wavenumber. We now examine the relevance of each formula, (2)-(4), using experimental data described below. We emphasize that the data of \cite{Nature} have been repeated in part \cite{JFMrot}, but a completely independent effort would be desirable.

%%%%%%%%%%%%%%%%%%%%%%%%%%%%%%%%%%%%%%%%%%
\section{Brief comments on the experimental data}
As stated above, our goals in the experiments of Ref.\ \cite{Nature} were to maximize the Rayleigh number attainable, and to place all high Rayleigh numbers within the turbulent regime, for obtaining robust scaling relations. As already stated, cryogenic helium gas was used as the working fluid for two main reasons: (1) It has the lowest kinematic viscosity of all known substances; and (2) by operating the experiment close to the critical point, the divergence of the specific heat $C_P$ means that $Ra \sim \alpha \rho^2 C_P$, where $\alpha$ is thermodynamically related to $C_P$, reaches extremely large values. Quoting from \cite{JLTP} where these factors were discussed in a bit more detail, we have: "For non-interacting gases, $\alpha = 1/T$, and so, low temperatures themselves have a particular advantage for buoyancy-driven flows. ... In summary, it is the combination $\alpha/\nu \kappa$ that determines
the Rayleigh number ..." There are two further advantages in using cryogenic helium: It is possible to stay closer to the Boussinesq approximation than in other fluids while attaining very high values of $Ra$, and one can achieve excellent thermal isolation. 

\vspace{0.3cm}
\begin{figure} [h]
%\begin{figure}
\includegraphics[width=10.5 cm]{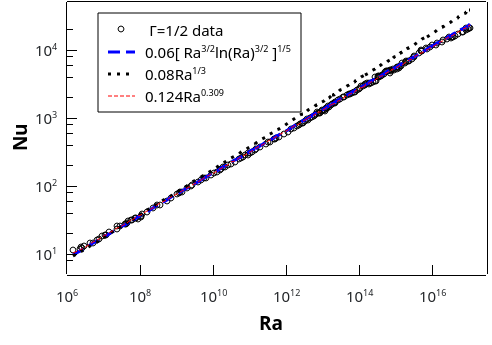}
\caption{Circles: data obtained with the 1-m tall, aspect ratio 1/2 experiment. Blue dashed line:  Herring's model, 
Eq.\ (4) with 
D=0.06, this being the one free parameter in the theory (see related text, especially Section 5). Red dashed line: Least squares fit to the raw data (Eq.\ (5)). Dotted line: $Nu \sim Ra^{1/3}$ for comparison. \label{fig1}}
\end{figure}   
\unskip
\vspace{0.3cm}

\section{Power law fits and Herring's formula }

Experimental data of \cite{Nature} suggest a power law with a scaling exponent slightly less than $1/3$. We showed in \cite{Nature} that the least squares fit to the data over the entire turbulent range ($10^6<Ra<10^{17}$) is
\begin{equation}
Nu=0.124Ra^{0.309}.
\end{equation}
This is indeed the simplest fit possible with two parameters: the amplitude and the slope of a linear fit of log($Nu$) vs log($Ra$). The fit (see the red dashed line in Figure 1) is excellent for the entire data. Note that there is no ambiguity in the data that the small difference of the exponent from 1/3 is real, perhaps to be regarded as some kind of intermittency correction; but there could also be other valid reasons for this slight departure. Indeed, numerical simulations at high $Ra$ in a slender convection cell \cite{iyer} suggest an exponent somewhat closer to 1/3. But it appears certainly far from the half-power in $Ra$. Indeed, at present, there is no convincing experimental support for Kraichnan's formula. 

%%%%%%%%%%%%%%%%%%%%%%%%%%%%%%%%%%%%%%%%%%

\begin{figure} [h]
\includegraphics[width=10.5 cm]{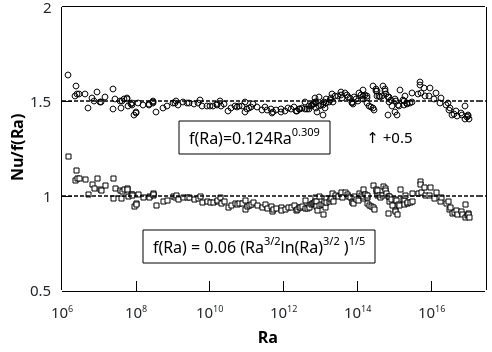}
\caption{$Nu$ normalized by Eqs.\ (4) and (5). Squares, $Nu$ normalized by Herring's formula (Eq.\ (4)). Circles, $Nu$ normalized by the least squares fit to the raw data, (Eq.\ (5)) with a vertical shift of +0.5 for clarity. For low $Ra < 10^{10}$, say, the slope is smaller than 0.309, reminiscent of the 2/7-th slope discussed in \cite{Castaing}. Our estimate is that the boundary layers in the apparatus of \cite{Nature} undergo transition at around $Ra = 10^{13}$. The last decade may have been influenced by non-constant Prandtl number (see next section) and/or moderate non-Boussinesq effects. These features are discussed at some length in \cite{jfm}. \label{fig2}}
\end{figure}

In Figure~\ref{fig1} we also show as the  blue dashed line Herring's prediction for turbulent convection between rigid boundaries, given by Eq.~(4), with the prefactor $D = 0.06$.
%\begin{adjustwidth}{-\extralength}{0cm}
%\begin{equation}
%Nu = 0.06  \lbrace Ra^{3/2}ln(Ra)^{3/2}\rbrace^{1/5}
%\end{equation}
%\end{adjustwidth}
The prefactor, the only free parameter in the expression, was obtained by fitting the function to the data. We note that the exponent 0.3 in Eq.~(4) is also applied to the ln($Ra$) term so that the effective exponent is slightly larger and, in fact, Herring's formula fits the data as well as the power-law fit, Eq.\ (5), as can be seen better in Figure~\ref{fig2}. We emphasize that, instead of two constants of the power-law, only the amplitude is an adjustable parameter in Herring's formula. 

In fact, Figure~\ref{fig2} shows the raw data normalized by both fits and it is clear that each of them is equally good and satisfactory overall. We also note that a considerable substructure to the data exists, which could suggest various changes in the flow, as hinted in the caption to Figure 2, while preserving the same global trend. We shall give in Section 5 a brief interpretation of the empirically determined prefactor in Eq.\ 4.  

\vspace{0.1cm}
\unskip
\vspace{0.3cm}

\section{Remarks on the ultimate state}

\begin{figure} [h]
\includegraphics[width=10.5 cm]{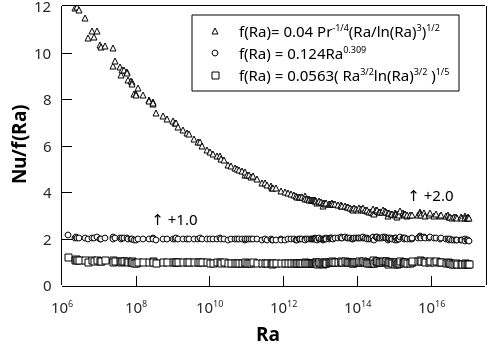}
\caption{$Nu$ normalized by Eqs.\ (4) and (5) as in Figure 2 but with compressed vertical scale to allow for normalization by Kraichnan's formula to appear on the same plot. The scale Squares, Nu normalized by Herring's formula (Eq.\ (4)). Circles, $Nu$ normalized by the least squares fit to the raw data, Eq.\ (5), with a vertical shift of +1.0 for clarity. Triangles, $Nu$ normalized according to Kraichnan's formula for high $Ra$ and moderate Prandtl number (Eq.\ (3) with $C=0.04$) with a vertical shift of +2.0 for clarity (see text). Here and in Figure 5, the dependence on Prandtl number is left implicit in the notation f($Ra$) as applied to Eq.\ (3). \label{fig3}}
\end{figure}

In Figure~\ref{fig3}, we compare the performance of the Kraichnan formula with those of the other two. To bring them all onto one plot, we set $C = 0.04$ in the Kraichnan formula and compressed the vertical scale by a factor of about 5. On this scale the bottom two fits mask the substructure apparent in Figure 2 and are almost perfect straight and horizontal lines, showing that both Eqs.\ (4) and (5) work very well from a global perspective. The Kraichnan formula is obviously far from being successful in reducing the data to a horizontal line, so it is clear that it has no global relevance unlike the other two. However, the curve appears to flatten for high $Ra$, tantalizingly suggesting a possible approach to the ultimate state. We shall consider this proposition below.

Figure 4 shows an enlarged plot of Nusselt number against $Ra$ for the last decade and a half or so. The best fit to those data is a power law exponent of 0.317, which is only minutely larger (by about 2.5\%) than the global exponent of 0.309 (and quite close to the finding in \cite{iyer}). No reasonable person would think that the slope is approaching a value of half. From a slightly different perspective, Figure 5 shows the high-$Ra$ end of the data of Figure 3; it is clear that the tendency to flatten, which might have been inferred from Figure 3, is an illusion caused by the compressed scale. It is not clear what functional form the normalized data in Figure 5 should take, but, if we fit a power law to the last two decades of $Ra$, it yields a power law with an exponent of $-0.033$.

An unstated argument sometimes adduced by the adherents of the ultimate state is to point out that the Prandtl number was not constant in the measurements of \cite{Nature} over the last two or so decades of $Ra$. This behavior was discussed at length in \cite{jfm} and is reproduced in Figure 6. 
%and is re Kraichnan's formula is intended to be applied for {\em constant and moderate} $Pr$, and is given here as
%\begin{adjustwidth}{-\extralength}{0cm}
%\begin{equation}
%Nu = 0.04\lbrace Pr^{-1/4} %Ra/log(Ra)^3\rbrace^{1/2},
%\end{equation}
%\end{adjustwidth}
%where the prefactor 0.04 is entirely arbitrary, chosen merely to move the formula in the same range as the experimental data. It is indeed the case that the range of $Ra$ where the data appear to approach a constant in terms of Kraichnan's normalization, $Pr$ varies by an order of magnitude in the experiment (as was discussed at some length in Ref.\ \cite{jfm}). This $Pr$-variation is shown in Figure~\ref{fig4} as a function of $Ra$. 
However, the variation of the Nusselt number on the Prandtl number is very weak for moderate Prandtl numbers in the range encountered here (see, e.g., \cite{lohse}). Thus, one cannot argue that the rise in Prandtl number is the reason why the data do not approach the half-power. Our conclusion could be different if the interpretation of the data has the benefit of a precise theory for how the heat transport depends on the Prandtl number.

%The variation in $Pr$, due to proximity to the critical point while obtaining very high $Ra$ can be described approximately, at high $Ra$, according to the fit
%\begin{adjustwidth}{-\extralength}{0cm}
%\begin{equation}
%Pr = 2 \times 10^{-8}Ra^{0.536}
%\end{equation}
%\end{adjustwidth}
%Substituting this into Kraichnan's formula gives a factor of $Ra^{-0.134}$ which reduces the factor $Ra^{1/2}$, and combined with division by $log(Ra)^{3/2}$, brings the effective exponent close to that of the experimental data. Our position is that unless we can obtain such high-$Ra$ data at constant Prandtl numbers, or have a full-fledged theory for the $Pr$-variation, or both, the case for Kraichnan's scenario cannot be made convincingly on the basis of the present data. 

\vspace{0.3cm}

%\unskip

\begin{figure} [h]
\includegraphics[width=10.5 cm]{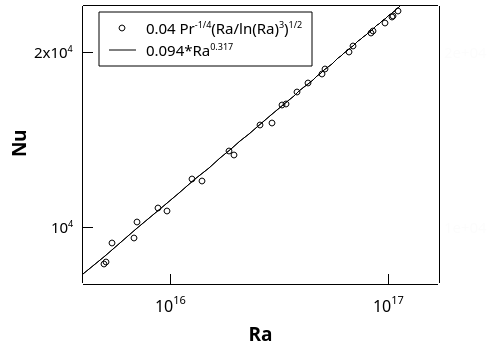}
\caption{$Nu$ vs $Ra$ for the highest decade and a half of $Ra$. Symbols: $Nu$ evaluated using Kraichnan's formula (Eq.\ (3) with $C=0.04$). Line: Least squares fit to the data giving a slope of 0.317. \label{fig4}}
\end{figure}   
\unskip

\vspace{0.3cm}
\begin{figure} [h]
\includegraphics[width=10.5 cm]{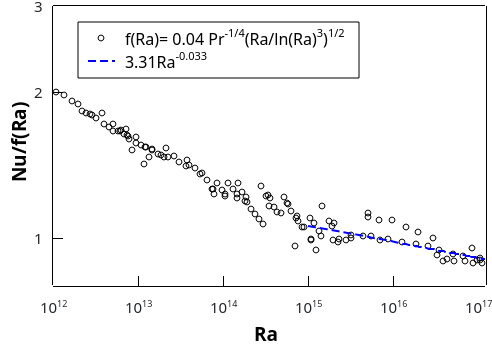}
\caption{$Nu$/f($Ra$) vs $Ra$. The data are normalized using f($Ra$) from Kraichnan's formula (Eq.\ (3) with C=0.04). Dashed line: least squares fit to the normalized data for $Ra > 10^{15}$. The slope over the last two decades is -0.033 indicating that the experimental data do not follow Kraichnan's formula. The prefactor 3.31 is dependent on the use of $C=0.04$ in Eq.\ (3). \label{fig5}}
\end{figure}   
%\unskip

\vspace{0.3cm}
\begin{figure} [h]
\includegraphics[width=10.5 cm]{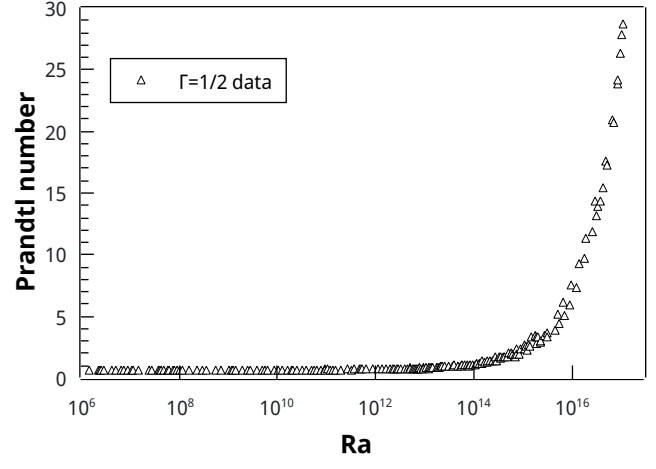}
\caption{Prandtl number vs Rayleigh number for the experiment of Ref.\ \cite{Nature}, showing a considerable variation at high $Ra$. This variation is an artifact of approaching the critical point of cryogenic helium gas to attain higher and higher Rayleigh numbers. This effect will be even more pronounced if one pushes for high $Ra$ in a smaller apparatus. \label{fig6}}
\end{figure}   
%\unskip
\vspace{0.1cm}

% Example of a figure that spans the whole page width. The same concept works for tables, too.
%%\begin{adjustwidth}{-\extralength}{0cm}
%\centering
%\includegraphics[width=15.5cm]{Definitions/logo-mdpi}
%\end{adjustwidth}
%\caption{This is a wide figure but actually a logo.\label{fig2}}
%\end{figure} 

%%%%%%%%%%%%%%%%%%%%%%%%%%%%%%%%%%%%%%%%%%
\section{Discussion}
The discussion here is mainly about two questions. The first question is why Herring's formula, derived by optimizing heat transport by single horizontal wavenumber in the slightly nonlinear supercritical regime, should work well for Rayleigh numbers up to $10^{17}$. Note that the critical Rayleigh number is 1708 for RBC, probably higher for this aspect ratio, say a few thousands. Yet the formula seems to apply for an extraordinary range of Rayleigh number range within which the flow is decidedly turbulent and the range of wavenumbers excited is continuous. 

One can perhaps say that this particular success of the Herring formula is a coincidence but that would be an unimaginative stance (given the large number of decades of $Ra$ over which the agreement occurs). So we will attempt to seek a possible meaning to the finding. The reason could perhaps be related to the later observation by Krishnamurti \& Howard \cite{Krishnamurti} of a large scale flow developing at high $Ra$, a phenomenon that has been observed widely in turbulent convection. Quoting from his 1966 paper \cite{Herring66}, Herring states (for free boundaries) that ``The physical picture of free boundary convective process predicted by the model is that of a large-scale motion dominating the central region between the conducting plates. This large-scale motion sweeps with it the temperature fluctuation field whose main variations occur in a thin boundary layer of vertical extent 1/Nu. The horizontal scale of both the dominant motion and the temperature fluctuation field is comparable to the distance between the conducting plates."  In actuality, this describes rather well the flow observed between rigid boundaries at high $Ra$. 

We now expand on this possibility. It is well known that the large structure in various turbulent flows resembles that in the supercritical nonlinear stage of evolution. A case in point is the wake behind a circular cylinder. Even at the highest Reynolds number measured \cite{rosh}, the large structure at high Reynolds number is very similar to the vortices shed just past the critical Reynolds number. We can speculate that the structure of the large scale remains unchanged but the {\it effective} amplitude of the transport coefficient changes to a new effective value as the small scale turbulent fluctuations set in. One may further speculate that convection has the same characteristic. If so, it stands to reason that the same functional form of the equation appropriate to the weakly nonlinear stage remains valid for the turbulent state, except that the prefactor will be different in the two states. This is renormalization in action, in a sense that is yet to be made precise. The only flow for which this program of renormalization has been carried out quantitatively is the isotropic and homogeneous flow generated by large scale forcing \cite{SV}. If this same program holds for convection, it must be possible formally to determine the effective diffusivity in the convection problem as well.

The second question concerns the ultimate state. We are aware that 1/2-power occurs in the presence of roughness \cite{topp}, when convection occurs in an open-ended tube \cite{arak}, when the fluid is heated by a body force such as radiation \cite{lepo}, etc. As long as the continuum equations hold and the boundary layers are intact, it appears to us that their importance to convection will not vanish, and that theories that have no place for viscosity and thermal diffusivity will miss an essential ingredient of the flow. To our knowledge, there is no compelling evidence to date, experimental or numerical, in favor of Kraichnan's formula. We have made that point explicit for the present data.  

%%%%%%%%%%%%%%%%%%%%%%%%%%%%%%%%%%%%%%%%%%
\section{Conclusions}

Jack Herring used a greatly truncated version of the complete Boussinesq equations containing only one horizontal wavenumber, and obtained a formula for calculating the Nusselt number in the weakly nonlinear supercritical state slightly past the critical Rayleigh number. The same formula works in the fully turbulent state extending over eleven orders of magnitude, if the numerical prefactor is suitably replaced. In this sense, the formula is less empirical in content than a single power-law which requires two constants to be determined from experiment. We have discussed why such a simplistic physical picture could work. We have also considered Kraichnan's asymptotic formula briefly and presented our reasoning why it does not hold for our experiment. We readily acknowledge that half-power law exists under different conditions of convection outside the standard RBC.

%%%%%%%%%%%%%%%%%%%%%%%%%%%%%%%%%%%%%%%%%%
%%%%%%%%%%%%%%%%%%%%%%%%%%%%%%%%%%%%%%%%%%
%\vspace{6pt} 

%%%%%%%%%%%%%%%%%%%%%%%%%%%%%%%%%%%%%%%%%%

%%%%%%%%%%%%%%%%%%%%%%%%%%%%%%%%%%%%%%%%%%

%\funding{The research shown here was was funded by the National Science Foundation grant number DMR-95-29609.}

%\dataavailability{The data used in the paper can be obtained in tabulated form by writing to one of the authors.} 
%We encourage all authors of articles published in MDPI journals to share their research data. In this section, please provide details regarding where data supporting reported results can be found, including links to publicly archived datasets analyzed or generated during the study. Where no new data were created, or where data is unavailable due to privacy or ethical re-strictions, a statement is still required. Suggested Data Availability Statements are available in section “MDPI Research Data Policies” at \url{https://www.mdpi.com/ethics}. 

\section{Acknowledgments}
%\acknowledgments{
We had the opportunity for interesting discussions with Jack Herring over the years and in various contexts, and one of the highlights was his model for high Rayleigh number convection. Jack was always graceful in his interactions. KRS remembers him as being the same from the very first time he met him in 1977 at NCAR to the last contact. Both of us are grateful that we knew him and have much pleasure in dedicating this short article to his memory. We wish to acknowledge the important role played by Ladislav Skrbek in making the measurements reported in \cite{Nature}.
%}

%\conflictsofinterest{The authors declare no conflict of interest.} 

%%%%%%%%%%%%%%%%%%%%%%%%%%%%%%%%%%%%%%%%%%

%%%%%%%%%%%%%%%%%%%%%%%%%%%%%%%%%%%%%%%%%%

%\reftitle{References}


%\bibliography{your_external_BibTeX_file}

%=====================================
% References, variant B: internal bibliography
%=====================================

%%%%%%%%%%%%%%%%%%%%%%%%%%%%%%%%%%%%%%%%%%
%% for journal Sci
%\reviewreports{\\
%Reviewer 1 comments and authors’ response\\
%Reviewer 2 comments and authors’ response\\
%Reviewer 3 comments and authors’ response
%}
%%%%%%%%%%%%%%%%%%%%%%%%%%%%%%%%%%%%%%%%%%
%\PublishersNote{}
%\end{adjustwidth}
\end{document}